\documentstyle[floats,aps,preprint]{revtex}
\begin{document}
\draft
\preprint{\vbox{\hbox{SOGANG-HEP 245/98} }}
\title{AdS${}_2$ and quantum stability in the CGHS model}
\author{Won Tae Kim\footnote{electronic address:wtkim@ccs.sogang.ac.kr}}
\address{Department of Physics and Basic Science Research Institute,\\
         Sogang University, C.P.O. Box 1142, Seoul 100-611, Korea}
\maketitle
\bigskip
\begin{abstract} 
The two-dimensional anti-de Sitter space(AdS$_2$) is constructed in terms of
the CGHS model. The geometric solutions are composed of the AdS vacuum
and the AdS black hole which are locally equivalent but
distinguishable by their mass.
The infalling classical fields do not play any role 
but the quantum back reaction is crucial 
in the formation of the AdS vacuum and AdS black hole.  
In the presence of the AdS black hole, 
there does not exist any radiation, which is consistent
with the constraint equations. Therefore the transition from 
the AdS black hole  to the AdS vacuum is impossible, and they are
quantum mechanically stable . We discuss the reason why 
the vanishing Hawking radiation appears in the AdS$_2$ black hole 
in contrast to asymptotically flat black holes.   
 
\end{abstract}

\bigskip
\newpage
\section{Introduction \hfill{ }}
Recently, there has been much attention to the
anti-de Sitter(AdS) spacetime
in connection with a calculation of the statistical entropy of 
the Ba$\tilde{{\rm n}}$ados, Teitelboim, and Zanelli(BTZ) black hole
\cite{btz,ct} which is related to the higher-dimensional black holes 
\cite{hyu}, and the 
supergravities on the boundary of AdS spacetime and conformal field theory
correspondence \cite{mal}.
It is now natural to study AdS spacetime which  may be essential to
resolve the quantum gravity puzzles.

On the other hand, 
black holes as geometric solutions of gravity theory are expected to
have Hawking temperatures \cite{haw} which means that one can detect
some radiation from the black hole through quantum fluctuations.
The calculation of Hawking radiation can be done on the
generic black hole backgrounds. 
At first sight, this Hawking radiation seems to be 
applied
to both asymptotically flat and nonflat geometries. 
In the former case, the Tolman
temperature as a local temperature is coincident with
the Hawking temperature in the asymptotic infinity, 
however, they are not compatible with each other in the
latter case, especially in AdS black holes
in two dimensions since the local temperature 
$T_{\rm local}=\frac{T_{\rm H}}
{\sqrt{g_{00}}}=\frac{\sqrt{M}}{2\pi \sqrt{r^2-Ml^2}}$
vanishes in the asymptotic infinity 
while the Hawking temperature is finite. Therefore,
it would be interesting
to study whether Hawking radiation appears or not in this black
hole. 
 
Some years ago, black hole evaporation and the back reaction
of the geometry has been studied by
Callan-Giddings-Harvey-Strominger(CGHS) \cite{cghs} and subsequently
by Russo-Susskind-Thorlacius(RST) \cite{rst} and in the many
literatures \cite{2ddg}. In two-dimensions, the quantum back
reaction of the geometry is more tractable compared to
the other higher dimensional cases and may solve various quantum
gravity problems \cite{ky}. 

In this paper, we study the quantum-mechanical 
generation of constant curvature
spacetime of AdS$_2$ in terms of quantum back reaction 
by using the CGHS(RST) model and obtain the AdS vacuum defined
as a lowest energy state of geometry and AdS black hole which is 
regarded as a massive state in Sect. II. 
Similarly to the CGHS model, we
take the large $N$ limit where $N$ is a number of conformal
matter fields in order to maintain the validity of semiclassical
approximations. The crucial difference from the CGHS solution is
that we shall assume the constant dilaton background instead of the 
linear dilaton or spacetime-dependent dilaton background.
In Sect. III, 
we shall calculate the Hawking radiation in this AdS$_2$
black hole and infer the Hawking temperature from it 
without resort to
the conventional definition of Hawking temperature. 
We find that the transition from 
the AdS black hole to the AdS vacuum is impossible, and they are
quantum mechanically stable.    
Finally some remarks and discussion will be given in Sect. IV. 

\section{Quantum Mechanical Generation of AdS$_2$ \hfill{ }}
In this section, we obtain the AdS black hole solution from the
CGHS model.
Let us now consider the two-dimensional low-energy string theory given by
\begin{equation}
\label{cg}
S_{{\rm{DG}}} = \frac{1}{2 \pi} \int\/d^{2}x\sqrt{-g}\, e^{-2\phi}
        \left[ R + 4(\nabla\phi)^2  + \frac{2}{l^2} \right],
\end{equation}
where $\phi$ is a dilaton field and the cosmological constant
is negative as $\Lambda=-\frac{1}{l^2}$.  
The action for the classical and quantum matter 
are written in the form of \cite{cghs,rst}
\begin{eqnarray}
S_{{\rm{Cl}}} &=&  \frac{1}{2 \pi} \int\/d^{2}x \sqrt{-g}\,
       \left[ - \frac12 \sum_{i=1}^N (\nabla f_i)^2 \right], 
\label{cm}\\
S_{{\rm{Qt}}} &=& \frac{\kappa}{2\pi}\int d^2x\/ \sqrt{-g}\left[
        -\frac14 R\frac{1}{\Box}R -\frac{\gamma}{2}\phi R \right],
\label{qm}
\end{eqnarray}
where the anomaly coefficient is given by $\kappa = \frac{N-24}{12}$
and for a good semiclassical approximation, we take the large number
of conformal matter fields. 
The parameter $\gamma$ is chosen associated with the models
such as $\gamma=0$ for the CGHS model and $\gamma=1$ for the RST model.  
The constant $\gamma$ will be in fact restricted in later.
The nonlocal Polyakov action \cite{pol} in Eq. (\ref{qm})
is written as, by introducing an
auxiliary field $\psi$ for later convenience,
\begin{equation}
S_{{\rm{Qt}}} = \frac{\kappa}{2\pi}\int d^2x\/ \sqrt{-g}\left[
        \frac14 R\psi-\frac{1}{16}(\nabla\psi)^2 -\frac{\gamma}{2}\phi
        R \right].
\label{qm-l}
\end{equation}
Then the effective total action is 
\begin{equation}
\label{total}
S_{\rm T} = S_{\rm DG} + S_{\rm M},
\end{equation}
where the matter part of the action 
is composed of two pieces of $S_{\rm M}=S_{\rm
Cl} + S_{\rm Qt}$. 

The equations of motion and the constraint
equations with respect to metric for the action (\ref{total}) are  
\begin{equation}
\label{eom}
G_{\mu\nu} = T_{\mu\nu}^{\rm M}
\end{equation} 
where
\begin{eqnarray}
& &G_{\mu\nu} = \frac{2\pi}{\sqrt{-g}}\frac{\delta S_{\rm DG}}
        {\delta g^{\mu\nu}} \nonumber \\
& & \qquad \/ =e^{-2\phi}\bigg[ 2\nabla_\mu\nabla_\nu\phi
        + 2g_{\mu\nu}\left((\nabla\phi)^2-\Box\phi -\frac{1}{2l^2}
        \right)\bigg],\label{cov:G} \\
& &T_{\mu\nu}^{\rm M} \ = -\frac{2\pi}{\sqrt{-g}}
        \frac{\delta S_{\rm M}}{\delta g^{\mu\nu}} \nonumber \\
& & \qquad \, =\frac12 \sum_{i=1}^N \left[\nabla_\mu f_i \nabla_\nu f_i -\frac{1}{2}(\nabla f_i)^2 \right] 
        + \frac{\kappa}{4}\left[\nabla_\mu\nabla_\nu\psi
        + \frac14\nabla_\mu\psi\nabla_\nu\psi
        - g_{\mu\nu}\left(\Box\psi+\frac18(\nabla\psi)^2
        \right)\right] \nonumber \\
& & \qquad \ \ \ -\frac{\gamma\kappa}{2}\bigg[
        \nabla_\mu\nabla_\nu\phi - g_{\mu\nu}\Box\phi\bigg].
                \label{cov:T} 
\end{eqnarray}
The remaining 
equations of motion with respect to dilaton, conformal matter fields,
and auxiliary field are given by respectively
\begin{eqnarray}
& & e^{-2\phi}\left[R+4\Box\phi-4(\nabla\phi)^2 +\frac{2}{l^2}
\right]=-\frac{\gamma\kappa}{4}  R ,\label{cov:phi}\\
& &\Box f_i =0, \label{cov:matt} \\
& &\Box \psi = -2R. \label{cov:psi}
\end{eqnarray}
The trace of Eq. (\ref{eom}) yields
\begin{equation}
\label{trace}
e^{-2\phi}\left[-2\Box\phi +4(\nabla\phi)^2 -\frac{2}{l^2}
\right]= \frac{\kappa}{2}R  +\frac{\gamma\kappa}{2}\Box \phi 
\end{equation}
where the right hand side of Eq. (\ref{trace}) are conformal
anomaly and local counter terms.
In the CGHS and RST models, the vacuum is 
a linear dilaton vacuum with a flat metric.
The dilaton charge 
$Q$ is nonzero, which is explicitly fixed through 
the condition of the asymptotic
flatness of the black geometry,
\begin{equation}
\phi(r)=-\frac12 Qr =-\frac{1}{\sqrt{2}l} r
\end{equation}
and it also receives quantum corrections \cite{cghs,rst}.

We now present new geometric solution called AdS$_2$ for the
constant dilaton background.             
Of course, the AdS$_2$ does not appear at the classical level, however
if we consider the quantum back reaction of the geometry, 
then the nontrivial geometry appears due to the conformal anomaly.
The dilaton field is now assumed to be a constant,
\begin{equation}
\label{dilaton}
\phi =\phi_0.
\end{equation}
Note that the constant dilaton solution is inconsistent at the classical
level of $\kappa=0$ as easily seen from Eq. (\ref{trace})
and it is only possible in the quantized theory.
This means that in quantum gravity there may appear new
kinds of geometries depending on the dilaton backgrounds.  
By using Eqs. (\ref{cov:phi}) and (\ref{trace}), the $\phi_0$
is chosen as
\begin{equation}
\label{condition}
\phi_0=-\frac{1}{2}\ln \left[\frac{\kappa(2-\gamma)}{4}  \right].
\end{equation}
From Eqs. (\ref{trace}) and (\ref{condition}) on the constant
dilaton background, the effective curvature scalar is obtained as 
\begin{equation}
R= -\frac{2}{l^2_{\rm eff}}
\end{equation} 
where $ l^2_{\rm eff}=\frac{l^2}{2-\gamma}$.
The parameter $\gamma$ is restricted to $\gamma < 2$ to obtain
the negative curvature scalar.
Then from Eq. (\ref{condition}), we can see the anomaly coefficient
should be $\kappa > 0 ~(N >24)$, which is automatically 
valid in the large $N$ 
limit. Especially, for $\gamma=1$ corresponding to the RST model  
\footnote{The nonflat solution for the RST model has been 
obtained in Ref.\cite{sol}.},
by adding ghost decoupling term \cite{str}, 
$\kappa$ can be shifted to $\frac{N}{12}$. 

  It is interesting to note that the constant curvature appears due to
the quantum back reaction of the metric. 
This interesting feature is essentially
on the basis of constant dilaton field. 
Most of cases, the linear dilaton 
vacuum and its time-dependence has been assumed in contrast to the present 
case. 

In the conformal gauge,
\begin{equation}
ds^2=- e^{2\rho(\sigma^+,\sigma^-)} d\sigma^+ d\sigma^-,
\end{equation}
the equations of motion and constraints (\ref{eom})
are given by
\begin{eqnarray}
&& \partial_+\partial_- \rho + \frac{1}{4l^2_{\rm eff}} e^{2\rho}=0,
            \label{dynamic} \\
&& \partial_+\partial_- f=0,\label{f} \\
&& T_{\pm \pm}^{M}=T_{\pm \pm}^{\rm cl}+T_{\pm \pm}^{\rm Qt}=0,\label{con} 
\end{eqnarray}
where $T_{\pm\pm}^{\rm cl}=\frac12\sum_{i=1}^N\left(\partial_\pm f_i\right)^2$,
 $T_{\pm \pm}^{\rm Qt}=   - \kappa\left[ \left(\partial_\pm\rho\right)^2- 
                \partial_\pm^2\rho\right] -\kappa t_\pm $, and 
       $t_{\pm}$ reflects the nonlocality of the conformal anomaly \cite{cghs}.
Solving the equations of motion  
in the conformal gauge yields \cite{dj,dfj,hen,str2}
\begin{eqnarray}
& & e^{2\rho}=\frac{M}{{\rm sinh}^2\left[\frac{\sqrt{M}(\sigma^+ -\sigma^-)}
                {2l_{\rm eff}} \right]},  \label{metric} \\
& & f_i =f_i^{(+)}(\sigma^+) +f_i^{(-)}(\sigma^-) \label{matsol}, 
\end{eqnarray}
where  
$M$ is an integration constant.
The AdS$_2$ vacuum is now defined by
\begin{equation}
\label{vacuum}
e^{2\rho}=\frac{4l^2_{\rm eff}}{(y^+ - y^-)^2}, 
\end{equation} 
and we assume that $M \ge 0$.
For $M \rightarrow 0$, the solution (\ref{metric}) exactly comes down to
the AdS$_2$ vacuum and the local geometries are equivalent in that
the curvature scalar is independent of the parameter $M$.
The parameter $M$ describes just only the existence of the horizon
of the geometry. 
Furthermore, the solution should be satisfied with 
the following constraint equations,
\begin{equation}
 \frac12\sum_{i=1}^N\left(\partial_{\pm} f_i^{(\pm)}\right)^2(\sigma^{\pm}) 
            -\frac{\kappa M}{4 l^2_{\rm eff}} -\kappa t_{\pm}(\sigma^{\pm}) =0.
\end{equation}
We shall assume there does not exist the classical flux, which is 
in fact
of no relevance to the formation and evaporation of black hole
as seen in Eq. (\ref{dynamic}).   
In fact, this classical flux of 
infalling matter fields cannot be connected 
with $M$ unless $t_{\pm}=0$. 
If we fix $t_{\pm}=0$ in the black hole
background, the Virasoro anomaly
appears in the constraints under the coordinate transformation
and the theory becomes inconsistent. 
On the other hand, 
if the classical matter fields exist, then
the boundary condition is just changed according to Eq. (\ref{con}). 
Either way, the classical matter field is not crucial in AdS$_2$ since
the constant curvature is independent of infalling matter energy
density. Therefore we simply set $T_{\pm \pm}^{\rm cl}=0$.

\section{Hawking radiation of AdS$_2$ black hole \hfill{ }}
Let us now study the Hawking radiation of this AdS$_2$ black 
hole.
The Hawking radiation in two dimensions 
is usually given by the anomalous transformation
of the energy-momentum tensor
\footnote{See Ref. \cite{gid} for extensive reviews}. 
This fact comes from the requirement
of the Virasoro anomaly free condition of the energy-momentum tensors.
Therefore $T_{--}^{Qt}$ should be a tensor without any anomaly under the
coordinate transformation. 
The Hawking radiation seems to be a global effect and
it is determined by the boundary effect given by the integration
constant $t_{-}$. Then it is given by \cite{gid,kl,cm}
\begin{eqnarray}
\label{sch}
-\kappa t_{-}(\sigma^-)&=&-\frac{\kappa}{2} \{y^-, \sigma^- \} \nonumber \\
                       &=&\frac{\kappa M}{4 l^2_{\rm eff}}
\end{eqnarray}
where $y^\pm =\frac{2l_{\rm eff}}{\sqrt{M}} {\rm tanh}
\frac{\sqrt{M}\sigma^\pm}{2l_{\rm eff}}$
and $\{y^-, \sigma^- \}$ is a Schwartzian derivative.
At first sight, 
the Hawking radiation seems to be a constant and it is compatible 
with the
Hawking temperature given by $T_{H}=\frac{\sqrt{M}}{2\pi l}$ since 
$-\kappa t_-=\kappa \pi^2 T_H^2$. However, this is not the case.   
The quantum-mechanical energy-momentum tensor
is defined 
as 
\begin{eqnarray}
\label{null}
h(\sigma^+,\sigma^-) &=&T_{--}^{\rm Qt} (\sigma^+,\sigma^-) \nonumber \\
                     &=& T_{--}^{\rm Bulk}+ T_{--}^{\rm boundary} 
                                  \nonumber \\     
                      &=& 0
\end{eqnarray}
since $T_{--}^{Qt}$ is composed of both bulk and boundary contribution
\begin{eqnarray}
T_{--}^{\rm Bulk}(\sigma^+,\sigma^-)& =&
- \kappa\left[ \left(\partial_-\rho\right)^2- 
 \partial_-^2\rho\right]\nonumber \\
 &=& -\frac{\kappa M}{4 l^2_{eff}} ,\label{bulk} \\  
T_{--}^{\rm boundary}(\sigma^+,\sigma^-)&=& 
-\kappa t_- \nonumber \\
&=& \frac{\kappa M}{4 l^2_{eff}} 
\label{boundary}
\end{eqnarray} 
respectively,
and they are 
exactly canceled out, which is consistent with Eq. (\ref{con}). 
The negative contribution of the bulk part Eq. (\ref{bulk}) is 
calculated by
the use of Eq. (\ref{metric}). It is interesting to note that
this part is constant, which is in contrast with
the asymptotically flat case, for instance, the CGHS black hole.
For the asymptotically spatial infinity, this null relation is
valid, and this means that there does not exist Hawking radiation
on the AdS black hole background. Therefore, any quantum
transition is impossible from the black hole state to the AdS$_2$ vacuum
through the Hawking radiation. 

At this stage,
it seems to be appropriate to compare Hawking radiation in the CGHS model 
with the present AdS$_2$ black hole.
For an asymptotically flat black hole of the CGHS model, the 
Hawking radiation is just given by \cite{cghs}
\begin{eqnarray}
h(\sigma^-)&=&T_{--}^{Qt}(\sigma^+,\sigma^-)
                  |_{\sigma^+ \rightarrow \infty}\nonumber \\
          &=& T_{--}^{\rm Bulk}|_{\sigma^+ \rightarrow \infty}+
            T_{--}^{\rm boundary}|_{\sigma^+ \rightarrow \infty} \nonumber \\ 
     &\approx& -\kappa t_-(\sigma^-)
\end{eqnarray}
where
\begin{eqnarray}
T_{--}^{\rm Bulk}(\sigma^+,\sigma^-)|_{\sigma^+ \rightarrow \infty}
   &=&- \kappa\left[ \left(\partial_-\rho\right)^2- 
 \partial_-^2\rho\right]|_{\sigma^+ \rightarrow \infty} \nonumber \\
&=&-\frac{1}{96l^2}\left[1-
              \frac{1}{\left(1+\sqrt{2}al e^{\frac{1}{\sqrt{2}l}(
               \sigma^- -\sigma^+ + \sigma^+_0   )}\right)^2}
       \right]~~\left( {\sigma^+ \rightarrow \infty}\right)\nonumber \\
&=& 0 \nonumber \\
T_{--}^{\rm boundary}(\sigma^+,\sigma^-)|_{\sigma^+ \rightarrow \infty}&=& 
-\kappa t_- \nonumber \\ 
&=&\frac{1}{96l^2}\left[1-
              \frac{1}{\left(1+\sqrt{2}al e^{\frac{1}{\sqrt{2}l}(
               \sigma^- )}\right)^2}
       \right]~~\left( {\sigma^+ \rightarrow \infty}\right)
\end{eqnarray}
and $a$ is a proportional to the infalling flux \cite{cghs}.
In the asymptotic null infinity $(\sigma^+ \rightarrow + \infty)$, 
Hawking radiation is only due to the
boundary term $t_-$  
since the bulk contribution
vanishes at the null infinity, 
while for the AdS$_2$ case both bulk and
boundary effects are simultaneously considered in the Hawking
radiation process since they are all constants. 

One may reconsider whether the infalling matter field affects the
formation and evaporation of AdS black hole or not. 
This problem can be studied by using the Jackiw-Teitelboim model \cite{jt},
\begin{equation}
S_{\rm JT}=\int d^2 x \sqrt{-g}\Phi \left[R +\frac{2}{l^2} \right]
\end{equation}
where $\Phi$ is an auxiliary field. 
From the beginning, we assume the AdS vacuum or AdS black hole background  
and consider the infalling conformal matter
field as in the CGHS model. Then the dynamical equation of motion
with respect to $\rho$ 
in the conformal gauge does not contain any information of 
matter fields similarly to our model. 

We now exhibit
some of equations different from the CGHS model $(\gamma=0)$ when we
consider the Polyakov induced gravity action (\ref{qm}) with the JT model,
\begin{eqnarray}
&&\partial_+\partial_- \Phi-\frac{1}{4l^2}e^{2\rho}(\frac{\kappa}{2}-\Phi)=0,\\
&&\partial_\pm^2\Phi - 2\partial_\pm \rho \partial_\pm \Phi=T_{\pm
  \pm}^{M}.
 \label{jtcon}
\end{eqnarray}
In this case, the solutions are given by
\begin{eqnarray}
\Phi^{-1}&=&-\frac{1}{M} {\rm tanh}\left[\frac{\sqrt{M}(\sigma^+ -\sigma^-)}
                {2l} \right]            , \nonumber \\
e^{2\rho}&=&\frac{M}{{\rm sinh}^2\left[\frac{\sqrt{M}(\sigma^+ -\sigma^-)}
                {2l} \right]}, 
\end{eqnarray}
where we simply assume $T^{\rm cl}_{\pm \pm}=0$.
Note that the constraint equations (\ref{jtcon}) should be Virasoro
anomaly free such that $T_{\pm\pm}^{\rm Qt}$ can be transformed as 
the primary
operator in conformal field theory,
\begin{equation}
T_{\pm\pm}^{\rm Qt}(\sigma^\pm)= \left(\frac{\partial_\pm y^\pm}{\partial_\pm
  \sigma^\pm}\right)^2T_{\pm\pm}^{\rm Qt}(y^\pm).
\end{equation}
In this case also, the bulk and
boundary effects contributes to the Hawking radiation 
and they are exactly canceled out as
\begin{eqnarray}
T_{--}^{\rm Bulk}(\sigma^+,\sigma^-)& =&
- \kappa\left[ \left(\partial_-\rho\right)^2- 
 \partial_-^2\rho\right] \nonumber \\
&=& -\frac{\kappa M}{4 l^2} ,\label{bulk2} \\  
T_{--}^{\rm boundary}(\sigma^+,\sigma^-)&=& 
-\kappa t_- \nonumber \\
&=& \frac{\kappa M}{4 l^2}. 
\label{boundary2}
\end{eqnarray} 
Therefore, as far as the energy-momentum
tensor of vacuum state vanishes $(T_{\pm\pm}^{\rm Qt}(y^\pm)=0 )$, the
radiation is impossible.
Note that in Refs. \cite{cm,lva}, the Hawking radiation is proportional
to the black hole mass due to Eq. (\ref{boundary2}). 
The crucial difference between them 
comes from the contribution of so called bulk part of
energy-momentum tensors. However, in Refs. \cite{cm,cad,lva}, the
models are in fact different in that the dilaton field is not constant,
and the back reaction of the geometry
may depend on the dilaton field with the metric, so the quantum-mechanical
energy-momentum
tensor may be different from that of our CGHS model in Sect. III. 

\section{Discussion \hfill{ }}

In this work, we have shown that the AdS black hole solution
is possible in the CGHS model. This is in fact realized in
the quantized theory by assuming the constant dilaton background.
Therefore, in the quantum level, there exist two kinds of 
black hole solutions, the CGHS and AdS solutions depending
on the dilaton charge.

However, one might think that the present 
result on null Hawking radiation of the two-dimensional
AdS black hole is doubtful. And it may be concluded that
the result may be
depend on details of some boundary conditions of AdS geometry.
So we now reconsider another way to clarify whether the Hawking
radiation comes out in this black hole or not. 
A massless scalar field as a test field is considered
on the AdS$_2$ black hole background, 
then the greybody
factor of this black hole can be evaluated through the wave
function of the test field. In this case, if we allow 
boundary condition compatible with the equation of motion of
scalar field,  
then the wave function at the horizon 
is decomposed into ingoing and
outgoing modes in the asymptotic infinity.
Remarkably, the amplitude of outgoing wave corresponding to the Hawking
radiation is zero for the massless scalar field, 
and the absorption coefficient is 1. 
This peculiar phenomena of vanishing Hawking radiation will
be discussed in detail in elsewhere \cite{kp} by comparing with
other models.  

On the other hand, there may be another reason why the
above trivial result comes out. It seems to be that
the two-dimensional AdS black hole is in fact locally
equivalent to the AdS vacuum in that the curvature scalar
is constant which is independent of the parameter $M$.
This parameter may be a coordinate artifact
since it can be removed by using the coordinate transformation as
$y^\pm =\frac{2l_{\rm eff}}{\sqrt{M}} {\rm tanh}
\frac{\sqrt{M}\sigma^\pm}{2l_{\rm eff}}$ where $y^\pm$ 
describes the vacuum geometry while $\sigma^\pm$ does the AdS
black hole. Therefore, one can think that the two-dimensional 
AdS black hole
and vacuum in some sense belong to the equivalent class. 
The only difference between them comes from the fact that
the parameter $M$ just globally describes the location of horizon
in the geometry. 
Furthermore, if the parameter $M$ turns out to be a coordinate
artifact, then it is meaningless to interpret it as a 
conserved quantity as a black hole mass.
The explicit ADM mass calculation on the background metric of AdS vacuum
\cite{cm}, gives interestingly vanishing ADM mass as far as we consider
the constant dilaton background in two dimensions, 
which is in contrast with the case of 
the three-dimensional BTZ black hole. 
This supports that the parameter $M$ is not a conserved 
mass but just a gauge artifact.
At this stage, one might again think that even though $M$ is trivial,
the horizon exists for the nonvanishing $M$, so there may be
thermal radiation similar to the Rindler space. In the Rindler
space, the accelerated observer detects the thermal radiation
which is related to the coordinate change, 
corresponding to the Schwartzian discussed in Eq. (\ref{sch})
and (\ref{boundary}) in our case. As pointed out in Sec. III,
in the AdS case, there exists another contribution, viz, the
bulk radiation Eq. (\ref{bulk}). Therefore, the net radiation
can be zero. 
 
The final point to be mentioned is 
that intuitively how come the AdS$_2$ black hole does not
radiate. The similar phenomena 
can be found in near horizon geometry of extremal charged black holes
of two-dimensional Maxwell-dilaton
gravity \cite{mny} or spherically symmetric reduced Reissner-Nordsrom solution
or most of D-brane solutions. Since the AdS geometry comes from the
extremal cases from the string theory point of view, 
so that Hawking radiation does not occur.
Therefore the present AdS$_2$ black hole might be an effective theory
of extremal black holes whose Hawking temperature is zero.

{\bf Acknowledgments}\\
WTK would like to thank S. Hyun, J. Lee, and S. J. Rey 
for exciting discussion on AdS$_2$ and 
J. J. Oh and J. H. Park for discussion on
greybody factor and also thank M. Cadoni for helpful discussion on 
Hawking radiation and S. Solodukhin 
for bringing his papers related
to AdS$_2$ to my attention. 
He is also very grateful to D. Vassilevich for
valuable comments.  
This work was supported by Korea Research Foundation, 
No. BSRI-1998-015-D00074 and the financial
support of Korea Research Foundation made in the program year of
1997."


\end{document}